\documentstyle[12pt]{article}
\newcommand{\be}{\begin{equation}}
\newcommand{\ee}{\end{equation}}
\newcommand{\bea}{\begin{eqnarray}}
\newcommand{\eea}{\end{eqnarray}}

\newcommand{\I}{\mbox{i}}
\newcommand{\D}{\mbox{d}}
\newcommand{\E}{\mbox{e}}

\begin{document}
\begin{titlepage}
\vskip 1cm
\begin{center}
\vfill
{\large\bf  THE COHERENCE OF PRIMORDIAL FLUCTUATIONS
            PRODUCED DURING INFLATION} 
\vskip 1cm
{\bf Claus Kiefer}
\vskip 0.4cm
 Fakult\"at f\"ur Physik, Universit\"at Freiburg,\\
  Hermann-Herder-Stra\ss e 3, D-79104 Freiburg, Germany.

\vskip 0.7cm
{\bf Julien Lesgourgues}
\vskip 0.4cm
 Lab. de Math\'ematiques et Physique Th\'eorique, UPRES A 6083 CNRS\\
  Universit\'e de Tours, Parc de Grandmont, F-37200 Tours, France.

\vskip 0.7cm
{\bf David Polarski}
\vskip 0.4cm
 Lab. de Math\'ematiques et Physique Th\'eorique, UPRES A 6083 CNRS\\
  Universit\'e de Tours, Parc de Grandmont, F-37200 Tours, France.\\
\vskip 0.3cm
 D\'epartement d'Astrophysique Relativiste et de Cosmologie,\\
  Observatoire de Paris-Meudon, F-92195 Meudon Cedex, France.

\vskip 0.7cm
{\bf Alexei A. Starobinsky}
\vskip 0.4cm
 Landau Institute for Theoretical Physics\\
  Kosygina St. 2, Moscow 117334, Russia.

\end{center}
\date{\today}
\vskip 2cm
\begin{center}
{\bf Abstract}
\end{center}
\begin{quote}
The behaviour of quantum metric perturbations produced during inflation 
is considered at the stage after the second Hubble radius crossing. It is 
shown that the classical correlation between amplitude and momentum of a 
perturbation mode, previously shown to emerge in the course of an effective 
quantum-to-classical transition, is maintained for a sufficiently long 
time, and we present the explicit form in which it takes place using the 
Wigner function. 
We further show with a simple diffraction experiment that 
quantum interference, non-expressible in terms of a classical stochastic 
description of the perturbations, is essentially suppressed. Rescattering of 
the perturbations leads to a comparatively slow decay of this correlation 
and to a complete stochastization of the system. 
\end{quote}

\end{titlepage}
 
\section{Introduction}
According to the inflationary scenario, all 
inhomogeneities in the Universe were produced quantum mechanically from 
quantum vacuum fluctuations of inflaton field(s) (effective scalar 
field(s) driving inflation) and gravitational field. Thus they are of genuine 
quantum gravitational origin and still observable today. 
The simplest variant predicts an approximately flat, or Harrison-Zeldovich, 
initial power spectrum for both scalar (density) perturbations 
\cite{Haw82} and gravitational waves (GW) \cite{al79}.
The corresponding fluctuations $\frac{\Delta T}{T}(\theta,\varphi)$ of the 
cosmic microwave background (CMB) were detected by the COBE large-angle, and 
a number of medium-angle, experiments. 
Clearly, all inhomogeneities measured presently appear to us as classical. 
So, a question of utmost importance is how and when does this effective 
quantum-to-classical transition occur? 

The amplitude of the observed fluctuations for a mode ${\bf k}$ is such 
that the condition $\langle n_{\bf k}\rangle\gg 1$,
where $n_{\bf k}$ is the particle number operator, 
is satisfied by a large 
margin (see \cite{PS96,LPS1} for exact conditions). However, this condition 
is not sufficient by itself to determine the nature of the 
quantum-to-classical transition, more information about the 
quantum state of each mode is needed. 
In particular, this transition is completely different for a coherent and 
a strongly squeezed state. 
The free non-relativistic particle at very late times provides yet another 
illustration of this dramatic difference \cite{KP}.
In the first case, the quantum mode can be approximated by a 
{\it deterministic} classical time- and space-dependent field 
(a Bose condensate).
But the generation of fluctuations at the inflationary stage through 
particle creation out of the vacuum (described by a Bogolyubov transformation 
in the Heisenberg representation) leads to a squeezed state. As emphasized 
recently \cite{PS96}, in the limit of extreme squeezing, a 
quantum mode can be approximated by a classical stochastic field with 
{\em random} (Gaussianly distributed) amplitude and {\em fixed} phase. 
This fundamental result can be extended to non-vacuum states \cite{LPS1}, 
with non-Gaussian distributed amplitude.
In this quantum to classical transition, it is the quantum correlation 
between the growing and decaying parts of the perturbations that gets lost 
(the decaying part may even be completely damped, e.g. by viscosity). 
As no interaction with the environment is needed here, in contrast to usual 
decoherence \cite{dec}, it was called ``decoherence without decoherence'' in 
\cite{PS96}.
The sensitivity of this process to a type of interaction with the environment 
was considered in \cite{KPS,KP}: when the decaying mode is negligible, as is 
the case in the high squeezing limit, the field amplitude basis is the 
classical ``pointer'' basis for interactions that are in field amplitude 
space. As this is the case for most interactions, the loss of quantum 
coherence between the growing and decaying parts of the perturbations is 
essentially independent of the details of these interactions.

These previous studies were mainly dealing with the behaviour of the
perturbations during the inflationary stage or after it, but still before 
the second Hubble radius crossing, i.e. for $\lambda\equiv \frac{a(t)}{k}\ge 
R_H$, where $a(t)$ is the scale factor of the FRW metric while 
$R_H=\frac{a}{\dot a}, c=1$ (see however eqs 54-56 and the discussion on 
p.388-389 in \cite{PS96}). Now we concentrate on the late stage of 
their evolution when $\lambda\ll R_H$.
The fixed phase of a perturbation mode mentioned above means that this part 
of the quantum coherence is {\it not} destroyed by decoherence up to the 
moment of the second Hubble radius crossing. It reveals itself in the form 
of a {\it classical} correlation between amplitude and momentum of the mode 
and is all that remains from the initial quantum coherence. 
Using the Wigner function, we show in Sec.2 that this correlation is 
maintained even for $\lambda\ll R_H$. Then, in Sec.3, we consider a simple 
diffraction experiment which shows again that quantum interference, 
non-expressible in terms of a classical stochastic description of the 
perturbations, is essentially suppressed. 
This long-existing classical correlation of cosmological perturbations 
was overlooked in many papers which discuss their decoherence and entropy 
(see e.g. \cite{BMP,EF}).
For scalar perturbations, this correlation leads to Sakharov oscillations 
in the matter transfer function and to oscillations in the dispersion values 
$C_l$ of the $\frac{\Delta T}{T}$ multipoles. Note, however, that these 
oscillations are of purely dynamical origin and do {\it not} discriminate 
between quantum and classical origin of the perturbations (see also the 
discussion in \cite{PLB95}). Of course, this correlation will not remain 
forever. Rescattering of perturbations and other processes as well  
finally lead to a complete loss of the initial quantum correlations, i.e. to a 
complete decoherence. As discussed in Sec.4, the characteristic time for the 
latter process depends crucially on the type and wavelength of the 
perturbations under consideration. 

\section{Classical correlation}
The time evolution of the perturbations in the regime $\lambda\gg R_H$
leads to an extreme squeezing which remains when $\lambda$ becomes smaller 
than $R_H$ (see e.g. \cite{PS96}).
As a result, the 
quantum coherence is expressible in classical stochastic terms: 
for a given ``realization" $y_{\bf k}$ of the fluctuation field, we  
have for its canonical momentum
 $p_{\bf k}\simeq \frac{g_{k2}}{f_{k1}}~y_{\bf k}\equiv p_{{\bf k},cl}$, 
the {\it classical} momentum for large squeezing, i.e.,
for $|r_k|\to \infty$.
Here, $f_k$, resp.$g_k$, is the amplitude, resp. momentum, field mode 
in the Heisenberg representation (we follow the conventions of \cite{PS96}), 
and the index $1$ ($2$) refers 
to its real (imaginary) part, the same convention is used for all complex 
quantities. The conformal time $\eta\equiv \int^t \frac{dt'}{a(t')}$ is used 
in the following.
This almost perfect classical correlation is best seen with the help of the 
Wigner function, and we show now what precise form it takes today for 
these perturbations deep {\em inside} the Hubble radius. 
Their modes $f_k$ can be written as 
\begin{eqnarray}
f_k & = & D_1~\sin (k\eta+\xi_k)+\I D_2~\cos(k\eta+\xi_k)
\label{DD}
\end{eqnarray} 
where $D_1,~D_2$ are real, and $\xi_k$ is some phase. 
Extreme squeezing, or equivalently the almost complete disappearance of the 
decaying mode during the evolution outside the Hubble radius, manifests 
itself in the ratio $\frac{D_2}{D_1}\propto \exp(-2r_k)$. 
This is of the order $10^{-100}$ or less for the largest cosmological scales! 
For an initial vacuum state the wave function is and remains Gaussian, 
while the Wigner function $W$ is positive definite. Then, for modes deep 
inside the Hubble radius, given by (\ref{DD}), the following Wigner function 
is obtained (we consider half of the phase-space)
\be
W(y_0,p_0;\eta)=\frac{1}{\pi}\E^{-\frac{y'^2}
{D_1^2}}~\E^{-\frac{(p'/k)^2}{D_2^2}}\ , ~~~~~~~~~~~~ 
\tan (k\eta + \xi_k)\gg {\rm e}^{-2r_k}~,\label{W}   
\ee
where the $(y'\frac{p'}{k})$-frame is rotating in the 
$(y_0\frac{p_0}{k})$-plane while the rotation is clockwise. The $y'$ axis 
makes an angle $\varphi_k$ with the $y_0$ axis that is
 given by $\varphi_k=\frac{\pi}{2}-k\eta-\xi_k$, the latter is just the 
squeezing angle.
The rotation velocity $\omega\equiv \frac{d\varphi_k}{dt}$, where $t$ is the 
cosmological time, is given by $\omega=2\pi~\lambda^{-1}_{phys}$, with 
$\lambda_{phys}\equiv a\frac{2\pi}{k}$.
Taking into account that $D_1\gg D_2$, it is clear that the Wigner function 
(\ref{W}) is concentrated along the $y'$ axis. 
This just corresponds to the classical random process 
$y=D_1~\sin(k\eta+\xi_k)~e_y$, where $e_y$ is classical Gaussian random 
variable with unit variance.  
The typical volume is a (rotating) highly elongated ellipse whose thickness 
is tremendously small and proportional to the amplitude of the decaying mode. 
It is in this in practice unobservable thickness, the variance of the 
quantity $p-p_{cl}$ when the ellipse is in horizontal position, that the 
quantum coherence not expressible in classical stochastic terms is 
``hidden''. We note that this typical volume remains 
constant during all the time evolution of the fluctuations inside as well as 
outside the Hubble radius \cite{LPS2}. It is clear that one is not allowed to 
average over the angle $\varphi_k$, as this would not reflect the remaining 
quantum coherence and the resulting fixed phase of the perturbations. For the 
largest cosmological scales it would correspond to averaging over times of 
the order of the age of the universe!  
Both terms in (\ref{DD}) will be of the same amplitude during a tiny time 
interval $\delta t$, per (half) oscillation, when they are both of order 
$\sim D_1~{\rm e}^{-2r_k}$. 
We have $\delta t \sim 10^{-80}~{\rm sec}$ or less for 
wavelengths on cosmological scales $\sim 100~h^{-1}{\rm Mpc}$!
%
%
Note that typical times for the loss of quantum coherence between these 
terms due to interaction with other fields are even less than $\delta t$ 
\cite{KP}. During this short time interval one has 
\be
W(y_0,p_0)\simeq \frac{1}{\pi}\E^{-\frac{(p_0/k)^2}{D_1^2}}~
    \E^{-\frac{y_0^2}{D_2^2}}~,\label{W1}
\ee
which still corresponds to the classical correlation between amplitude and 
momentum. In conclusion, we still have zero measure ``trajectories'' in phase 
space satisfying $p=p_{cl}(y)$ with an accuracy well beyond 
observational capabilities.

\section{A diffraction experiment}
We shall now show with a concrete diffraction (gedanken) experiment that 
quantum interferences are quasi-classical up to an accuracy well beyond 
observational capabilities. 
We consider the following experiment: at time $\eta_1$, 
an apparatus rejects all fluctuations with amplitudes outside the range 
$[-\Delta, \Delta]$. The fluctuations that got through are then observed at 
some later time $\eta_2$. 
Our experiment corresponds physically to a fixed 
``slit'' of size $2\delta$ in $\phi$-space and  
to a slit of time-dependent size $2a\delta\equiv 2\Delta$ in $y$-space where 
$a(\eta)$ is the scale factor. 
%
%
For our sake it is sufficient to consider the real part, $y_{{\bf k}1}$ and
we further introduce the simplified notation 
$y_1(\eta_1)\equiv x_1,~y_1(\eta_2)\equiv x_2$.

We would like to show that the observed ``pattern'' 
corresponds to the predictions of a classical stochastic process to 
tremendous high accuracy.
The outcome of our experiment is encoded in the quantity $I(\Delta)$, or 
better $|I(\Delta)|^2$ which gives the probability distribution at time 
$\eta_2$
\be
I(\Delta)\equiv \int_{-\Delta}^{\Delta} \D x_1 K( x_2,\eta_2; x_1,\eta_1 ) 
\Psi(x_1,\eta_1)~,\label{I}
\ee
where the propagator $K( x_2,\eta_2; x_1,\eta_1 )$ gives the 
probability amplitude to go from $x_1$ at time $\eta_1$ to 
$x_2$ at time $\eta_2$ \cite{Feyn}. 
%
%
%
We perform our diffraction experiment with fluctuations on cosmological 
scales deep inside the Hubble radius, $\frac{k}{a}\gg H$. 
Hence the wave function $\Psi(x_1,\eta_1)$ in (\ref{I}) corresponds to a 
highly WKB state with $\Psi(x_1,\eta_1)=R(x_1,\eta_1)~
\E^{\frac{i}{\hbar}S_{cl}(x_1,\eta_1)}$ \cite{PS96,KP}. 
After some calculation the following result is obtained
\be
K( x_2,\eta_2; x_1,\eta_1 )\Psi(x_1,\eta_1)\propto R(x_1,\eta_1)~ 
\E^{\frac{i}{\hbar}\frac{k}{\sin k\Delta \eta}\left (\frac{f_1(\eta_2)}
{f_1(\eta_1)}x_1^2 - 2 x_2~x_1 \right )}~.\label{int}
\ee
There is a range of extremely narrow slits, 
of size $2\delta$ in $\phi$-space or equivalently $2\Delta$ in $y$-space, for 
which the real part of the wave function 
can be taken to be constant across the slit, while the complex phase cannot, 
and we assume our slit satisfies this condition. Hence the details of 
the wave function are irrelevant provided the state is of the WKB type. 
Therefore, $I(\Delta)$, see (\ref{I}), is given by an expression of the type 
(\cite{Grad}, 2.549.3-4)
\be
I(\Delta)\propto \sqrt{\frac{\pi}{2d}} 
\E^{-i \frac{b^2}{d}} \Bigl \lbrace C(\frac{d\Delta + b}{\sqrt{d}}) + 
C(\frac{d\Delta - b}{\sqrt{d}})+ 
\I\bigl \lbrack S(\frac{d\Delta + b}{\sqrt{d}}) + 
S(\frac{d\Delta - b}{\sqrt{d}}) \bigr \rbrack  \Bigr \rbrace~,\label{CS}
\ee
where
$d\equiv \frac{k}{\hbar~\sin k \Delta \eta}~\frac{f_1(\eta_2)}{f_1(\eta_1)},~
b\equiv -\frac{k}{\hbar~\sin k \Delta \eta}~x_2~$.
The result is thus expressed in terms of the Fresnel integrals $C(x),~S(x)$.
%
%
Even a slit of very small constant size $2\delta$ in $\phi$-space will 
eventually satisfy $2\Delta \gg 1$ at time $\eta_1$ due to the inflationary 
growth of the scale factor. 
An essential property here of the Fresnel integrals is that 
they both tend to $\frac{1}{2}$ for $x\to\infty$, they are close to this 
asymptotic value already at $x\simeq 10$ (see e.g. \cite{Bruhat}).
Hence, by inspection of (\ref{CS}), it is seen that for 
$|d|~\Delta \gg 1$ the expression (\ref{CS}) rapidly becomes zero when
$|b|>|d| \Delta$, within a range (in $y$-space) which is exceedingly small 
compared to $|d|~\Delta\gg 1$.
This means that we get a sharp pattern with $|\Psi (x_2)|^2\simeq |\Psi (0)|^2$
in the range 
\be
-\frac{f_1(\eta_2)}{f_1(\eta_1)}\Delta \leq x_2 \leq 
 \frac{f_1(\eta_2)}{f_1(\eta_1)}\Delta~.
\ee 
This is precisely the result one would expect with a classical stochastic 
process.
  
\section{Loss of remaining coherence}

As we have seen, there exists a long period in the evolution of a 
perturbation after 
the second Hubble radius crossing when it oscillates with a fixed phase and 
occupies a very small volume in phase-space. Certainly, this cannot proceed 
forever, and eventually we may expect total stochastization and total loss 
of the initial coherence (correlation) to take place. How fast it occurs, if 
it occurs at all, depends on the concrete system under consideration. 
\vskip 10pt
\noindent
{\bf a) Gravitational waves}: The first process which can lead 
to stochastization of the phase is graviton-graviton scattering. As was 
discussed in \cite{KPS}, the pointer 
observable (the amplitude in this case) does not commute with the 
corresponding interaction Hamiltonian. However, a simple estimate shows that 
this interaction is ineffective even for wavelengths crossing the Hubble 
radius for the first time towards the end of inflation.
Another process seems to be more effective: generation of a secondary GW 
background by matter after the second Hubble radius crossing. This process 
does not respect the phase of the primordial background and produces GW with 
a stochastic, uniformly 
distributed, phase. To screen the fixed phase of the primordial background, 
the spectral density $\frac{d\epsilon_g}{d\ln \nu}$ of the secondary 
background should be larger than the primordial one which is 
$< 10^{-14}\rho_{crit}$, with $\rho_{crit}$ being
 the critical energy density. 
While there exist astrophysical sources which may produce such a large 
secondary 
background with frequencies $\nu > 10^{-4}$Hz, a large secondary background 
at smaller frequencies requires exotic sources in the early Universe like, 
for example, cosmic strings (see e.g. \cite{Al}).
\vskip 10pt
\noindent
{\bf b) Scalar perturbations}: Scalar perturbations with present 
scale $\lambda > R_{eq}\sim 15~h^{-2}~
{\rm Mpc}$ crossed the Hubble radius last during the matter-dominated stage.
In contrast to GW, they do not oscillate. For them, the classical correlation 
between amplitude and momentum discussed above simply means that (in the 
linear approximation) the velocity potential $\Psi_{vel}$ is proportional 
to the gravitational potential: $\Psi_{vel}=\Phi~t$ (in other words, velocity 
and acceleration are parallel). Non-linear effects change the relation 
between both potentials without destroying it, until shell-crossing occurs. 
After that, the motion of matter 
becomes multistreamed and cannot be described anymore with one-fluid 
hydrodynamics. This results in the growth of the phase-space occupied by the 
perturbations. It is natural to expect that the initial phase relation will 
be destroyed in the high density regions with $\frac{\delta\rho}
{\rho}\gg 1$ where gravitational relaxation took place. 
Hence, the characteristic time $\tau$ for the loss of the remaining coherence 
is $\tau\sim t_k~\Phi^{-\frac{3}{2}}$, where $\Phi\equiv \sqrt{k^3\langle 
\Phi_k^2\rangle}\sim 10^{-5}$ 
is the r.m.s. amplitude of the initial gravitational potential and $t_k$ 
is the second Hubble radius crossing time.

Scalar perturbations with present scale $\lambda < 15h^{-2}~{\rm Mpc}$ passed 
through the stage of acoustic oscillations in the past. At that stage, 
Thompson scattering of photons by electrons leads to a dissipative process, 
the so-called Silk damping. However, it is not clear yet if this process 
destroys the phase correlation quicker than it damps the acoustic 
oscillations themselves. In particular, though numerical simulations clearly 
show that both the multipoles $C_l$ and their acoustic peaks are strongly 
damped for $l>1000$, it remains yet to be shown that the latter is damped 
quicker. Hence, even scalar perturbations on scales 
$\lambda\ll 15h^{-2}~{\rm Mpc}$ nowadays might well still keep the 
correlation in question.
\vskip 10pt
\noindent
{\bf c) Reheating after inflation}: An instructive example of 
formation and decay of classical correlations in 
a quantum system is given by preheating -- rapid creation of Bose particles 
by an oscillating inflaton field in the regime of broad parametric resonance 
\cite{KLS94}. Here, the practical implementation of the idea \cite{PS96} that 
neglection, for each mode, of an exponentially decaying part as compared 
to the exponentially growing part, is sufficient for effective 
quantum-to-classical transition has led to the solution of the corresponding 
classical inhomogeneous wave equation with stochastic initial conditions 
\cite{KT} as 
a way to go beyond the Hartree-Fock approximation (or its variants like the 
$\frac{1}{N}$ expansion). In the Hartree-Fock approximation, numerical 
calculations show the formation of a fixed phase for a given ${\bf k}$ mode 
when the number of created particles $\langle n_{\bf k}\rangle$ becomes large 
\cite{BdV} (this is reflected in particular by the fact that 
$\langle \phi_{\bf k}^2 \rangle$ oscillates much more rapidly than the 
external 
inflaton field), and no decay of this correlation is seen at late times. 
On the other hand, higher loop effects like rescattering of created particles 
which are automatically taken into account in the stochastic numerical 
simulations finally lead to complete loss of the correlation after a 
sufficiently large number of oscillations \cite{KT97}.

\vspace{1cm}
\noindent
{\bf Acknowledgements}
\par\noindent
A.S. acknowledges financial support by the Russian Foundation for Basic 
Research, grant 96-02-17591, by the Russian research project 
``Cosmomicrophysics'' and by the INTAS grant 93-493-ext.

\end{document}